\documentclass[conference]{IEEEtran}
\IEEEoverridecommandlockouts
\usepackage{cite}
\usepackage{amsmath,amssymb,amsfonts}
\usepackage{algorithmic}
\usepackage{graphicx}
\usepackage{textcomp}
\usepackage{xcolor}
\usepackage{setspace}
\def\BibTeX{{\rm B\kern-.05em{\sc i\kern-.025em b}\kern-.08em
		T\kern-.1667em\lower.7ex\hbox{E}\kern-.125emX}}
\usepackage{subfigure}
\newtheorem{my_theorem}{Theorem}

\newtheorem{my_lemma}{Lemma}

\newtheorem{my_proposition}{Proposition}

\title{Performance Analysis of Relay-Assisted	OWC  Over  Foggy 	Channel with Pointing Error}
\author{
	\IEEEauthorblockN{Tejas Nimish Shah, Ziyaur Rahman, and S.M. Zafaruddin}\\
	\IEEEauthorblockA{ Deptt. of Electrical and Electronics Engineering, 
		BITS Pilani, Pilani-333031, Rajasthan, India.\\ Email: \{f20170024, p20170416, syed.zafaruddin\}@pilani.bits-pilani.ac.in}
		}

\begin{document}
	\maketitle
	\begin{abstract}
Signal fading due to atmospheric channel impairments and pointing error is a major bottleneck for the performance of optical wireless communication (OWC).
In this paper,  we consider an amplify-and-forward (AF) optical relaying to enhance the performance of the OWC system with a negligible line-of-sight (LOS) link under the combined effect of fog and pointing error.  We analyze the end-to-end performance of the relay-assisted  system, which consists of complicated probability distribution functions.  We  derive analytical expressions of the outage probability,  average signal-to-noise ratio (SNR), and ergodic rate in terms of OWC  system parameters.  We also develop an exact integral-form expression of these performance metrics using the half-harmonic mean of individual SNRs  to validate the tightness of the derived analytical expressions.  The numerical and simulation analysis shows that the proposed dual-hop relaying has significant  performance  improvement when comparing to the direct transmission over considered channel impairments. Compared to the direct transmission, the relay-assisted system requires almost $30$ times less transmission power to achieve the same outage probability. The considered system  also provides a significant gain in the average SNR and ergodic rate for practical scenarios of OWC deployment. 

\end{abstract}
	\section{Introduction}
Optical wireless communication (OWC) is an emerging broadband technology that transmits data through an unguided atmospheric channel  in the unlicensed optical spectrum \cite{Khalighi2014, Kedar2004}.  However, signal transmission at  small wavelengths encounters  different channel impairments such as  atmospheric turbulence (due to the scintillation effect of light propagation),  pointing error  (caused by the  misalignment between the transmitter and receiver), and fog. The  pointing error  caused by the dynamic wind loads, weak earthquakes and thermal expansion  has a statistical effect on  the signal quality and presents a major challenge  in the  OWC deployment \cite{Farid2007, Vavoulas2012}. On the other hand, the impact of foggy conditions on OWC systems depends on the intensity of fog, which ranges between light, medium and dense.  It is noted that turbulence and fog may not co-exist since both are  inversely correlated with each other. This  allows us to investigate the effect of  turbulence and foggy conditions separately on the OWC  performance. 


The use of relaying to improve the performance of OWC systems has  been extensively studied under the effect of turbulence and/or pointing errors \cite{assym_rf_fso2015, series_hybrid_m_channel2015, dual_hop_rf_fso_turb2016, hybrid_outage2018, multi_rf_fso2017,multi_source_relay2014, Safari2008, dual_hop_turb2017,Yang2014_relay, multi_hop_turb2015,multi_fso2013,parallel_multi_fso2016,parallel_fso2015}. Hybrid RF/OWC systems, where relays act as an interface between RF and optical links, have been studied in  \cite{assym_rf_fso2015, series_hybrid_m_channel2015, dual_hop_rf_fso_turb2016, hybrid_outage2018, multi_rf_fso2017,multi_source_relay2014}. Considering a dual hop transmission with a single-relay and ignoring the direct transmission, the authors in  \cite{assym_rf_fso2015, series_hybrid_m_channel2015, dual_hop_rf_fso_turb2016} have analyzed the OWC performance under the  turbulence and pointing error. The use of multiple relays in parallel with active relay-selection is shown to improve the reliability of an OWC system in \cite{hybrid_outage2018, multi_rf_fso2017}. A multi-user dual-hop system using a multi-aperture RF relay over mixed links is investigated in  \cite{multi_source_relay2014}.

Recently, relaying for all-optical OWC systems has gained considerable research interests \cite{Safari2008, dual_hop_turb2017,Yang2014_relay, multi_hop_turb2015,multi_fso2013,parallel_multi_fso2016,parallel_fso2015}. In the seminal work \cite{Safari2008}, both multi-hop and cooperative  relaying coupled, with amplify-and-forward and decode-and-forward modes has been considered only for turbulence channel. The authors in  \cite{dual_hop_turb2017, Yang2014_relay} provided asymptotic expressions for  outage probability, average bit-error-rate (BER) and ergodic capacity under turbulence and  pointing errors by considering a single amplify-and-forward (AF) relay with fixed and variable gain. Multi-hop relaying with AF and decode-and-forward (DF) protocol, under the Gamma-Gamma turbulence and pointing error,  has been considered in \cite{multi_hop_turb2015, multi_fso2013}. The authors in  \cite{parallel_multi_fso2016} have considered inter-relay cooperation to  boost the reliability  of an OWC system. In a line-of sight communication, a single relay can be used to enhance the direct transmission both the received signals at the destination, as shown in \cite{parallel_fso2015}.

In the aforementioned and related research, the statistical  effect of foggy channels has not been considered.   Traditionally, signal attenuation due to the fog was quantified using a visibility range, less in light fog and more in dense fog \cite{Kim2001,Kedar2003}. However, recent studies confirm that signal attenuation due to fog is not deterministic but Gamma distributed \cite{Awan2008,Esmail2016_ICC,Esmail2016_Photonics, Berenguer2018}. Following this direction, the authors in \cite{Esmail2017_Access, Rahman2020} have analyzed the impact of fog on OWC using  various performance measures such as average signal-to-noise ratio (SNR), ergodic rate, outage probability, and BER. They have shown that  OWC performance is significantly limited in dense fog, but can provide  acceptable performance in light fog over short links. However, combining the effect of pointing error with fog shows high degradation in performance even in light foggy conditions\cite{rahman2020cl, Esmail2017_Photonics}. The authors in \cite{Esmail2017_Photonics} have considered three techniques to  mitigate this effect, those being multi-hop relay systems using DF, active laser selection and parallel RF/OWC links.  Laser selection and hybrid transmission techniques require feedback from the receiver to the transmitter, thereby increasing the overhead. Multi-hop relaying on the other hand requires channel state information (CSI) at each relay, to decode the signal, which can be hard in practice. It is noted that there are no analyses of average SNR, ergodic rate, and diversity order for the performance improvement of a relay-assisted OWC system under the effect of fog and pointing errors.

In this paper,  we  analyze the end-to-end performance of a relay-assisted OWC system under the combined effect of  fog and pointing error. We consider a single optical-relay with AF relaying, assuming no direct transmission to the destination. Using an upper bound approximation of the end-to-end SNR,   we derive closed-form expressions for the outage probability,  average SNR, and  ergodic rate of the relay-assisted OWC system. It should be noted that it requires novel approaches to analyze the end-to-end system since distribution functions of individual links  consists of  incomplete gamma functions and exponential integrals. The analysis show that the proposed dual-hop relaying improves the performance of the OWC system significantly.  We also develop an integral-form expression for the probability density function (PDF) of the end-to-end SNR  to validate the tightness of the derived analytical expressions.   Numerical evaluation is presented to validate the results of the analytical formulae, demonstrating the improvement in  performance compared to the direct transmission.

\section{System Model}\label{sec:system_model}
We consider an OWC system using intensity modulation/direct detection (IM/DD). It consists of a single-aperture transceiver system, with a negligible line-of-sight (LOS) link, under fog and pointing error.
The signal received at the receiver aperture, $y$, is given as  
\begin{eqnarray}
	y = h_{f} h_{p}Rx + w,
	\label{received signal}
\end{eqnarray} 
where $x$ is the transmitted signal, $R$ represents the detector responsivity (in amperes per watt), and $w$ represents an Additive White Gaussian Noise (AWGN) with variance $\sigma^2_{w}$. The terms $h_{f}$ and $h_{p}$  are the random states of the foggy channel and pointing error, respectively.

We define $\gamma=\gamma_0 |h|^2$ as the SNR, where $h = h_ph_f$, $\gamma_0=2P^2_tR^2/\sigma^2_w $,  and $P_t$ is the average optical transmitted power.
Denoting  the Gamma function as $\Gamma(x)=\int\limits_{0}^{\infty}t^{x-1} e^{-t}dt$ and the incomplete Gamma function as $\Gamma(a,t) =\int_{t}^{\infty}s^{a-1}e^{-s}ds$, the PDF of SNR for the OWC system can be easily derived using the distribution of $|h|^2$  \cite{Esmail2017_Photonics}:
\begin{eqnarray}
	&	{f_\gamma}(\gamma)=\frac{z^k \rho^2   }{2 m^k\Gamma(k)A_0^{\rho^2} \sqrt{\gamma\gamma_0}}\left(\sqrt{\frac{\gamma}{\gamma_0}}\right)^{\rho^2-1} \nonumber \\& \times [\Gamma(k)-\Gamma(k,m \ln(A_0/\sqrt{\gamma/\gamma_0}))],~~ \gamma\leq A_0^2 \gamma_0
		\label{snr_pdf}
\end{eqnarray}
where, $k>0$ is the shape parameter, $\beta>0$ is the scale parameter, $z=4.343/\beta d$, $d$ (in  \mbox{km}) the transmission link length between source and destination, and $m=z-\rho^2$. The parameter  $\rho=\omega_{z_{eq}}/2\sigma_s$ is the ratio between the equivalent beam radius and the standard deviation of the pointing error displacement at the receiver \cite{Esmail2017_Photonics}. The parameter $A_0$ is the fraction of collected power given as $A_0=(\mbox{erf}(\upsilon))^2$ where $\upsilon=\sqrt{\pi/2}\ a/\omega_z$, and $a$ is the receiver aperture radius. Since  experimental data for pointing error  parameters $\rho$ and $A_0$ is available only for few link ranges,   we obtain  a simple expression to determine these parameters for  various link distances for numerical analysis in  Section IV.

Denoting $E_n(a,r)=\int\limits_{1}^{\infty}\frac{e^{-rt}}{t^a}dt$ as the two-argument exponential integral,  we can obtain the cumulative distribution function (CDF) of foggy channel with pointing error \cite{Esmail2017_Photonics} :
\begin{eqnarray}
			&{F_\gamma}(\gamma)=\frac{z^k}{m^k ({A_0}/\sqrt{\gamma/\gamma_0})^{\rho^2}}+\frac{z^k}{m^k\Gamma(k)}\big[-e^{-\rho^2\ln ({A_0}/\sqrt{\gamma/\gamma_0})} \nonumber \\ &\Gamma(k,m\ln ({A_0}/\sqrt{\gamma/\gamma_0}))+((\rho^2+m)\ln ({A_0}/\sqrt{\gamma/\gamma_0}))^{-1} \nonumber \\& (m\ln({A_0}/\sqrt{\gamma/\gamma_0}))^k\big(e^{-(\rho^2+m)\ln ({A_0}/\sqrt{\gamma/\gamma_0})}+\nonumber \\&(k-1)E_n(2-k,(\rho^2+m)\ln ({A_0}/\sqrt{\gamma/\gamma_0}))\big)\big]
		\label{snr_cdf}
\end{eqnarray}
For the case of relayed transmission, the expressions for signals received at the relay and destination are:
\begin{eqnarray}
	y_r = h_{f1} h_{p1}Rx + w_1
	\label{relay signal}
\end{eqnarray}
\begin{eqnarray}
	y_d = h_{f2} h_{p2}Gy_r + w_2
	\label{destination signal}
\end{eqnarray}
where $h_{f1}$, $h_{p1}$ and $h_{f2}$, $h_{p2}$ are random fog and pointing error states between source-relay and relay-destination, respectively, each having $w_1$ and $w_2$ as AWGNs. $h_{1} = h_{f1}h_{p1}$ is the combined channel between source and relay, and $G = \sqrt{\frac{P_t}{h_{1}^2P_t^2R^2 + \sigma^2_w}}$ is the antenna gain. Here, $x$ and $y_r$ are transmitted from the source and relay respectively.

Taking $P_t = P_r$,  we get the instantaneous SNRs of signals received at the relay and receiver as $\gamma_1 = \gamma_{0}h_{1}^2$ and $\gamma_2 = \gamma_{0}h_{2}^2$, respectively. Using these, the expression for the end-to-end SNR is given as \cite{Yang2014_relay}:
\begin{eqnarray}
		\gamma_{e2e}=\frac{\gamma_1\gamma_2}{\gamma_1+\gamma_2+1}
		\label{inst_snr}
\end{eqnarray}
To derive tractable analytical  expressions using the well known half harmonic mean expression, we ignore the $1$ (since the PDF value is small for $\gamma<1$) in the numerator to get:
\begin{eqnarray}
	\gamma_{e2e}=\frac{\gamma_1\gamma_2}{\gamma_1+\gamma_2} \leq \min(\gamma_1, \gamma_2)
	\label{harmonic mean}
\end{eqnarray}

 We assume similar foggy conditions (i.e., same $k$ and $\beta$) for both links. This assumption is practical since foggy conditions do not change over distances of a few kilometers, over which a majority of OWC systems work. However, we get different values for $\rho$ and $A_0$ depending on the individual link lengths.
We also consider that  the relay is situated at the mid-way  between the source and destination since this results into optimum performance as demonstrated in numerical section. This causes the  parameters $A_0, w_z$, $\rho, m$ and $z$ same for both links (i.e., a symmetric condition) facilitating insightful analytical expressions.   

\section{Performance Analysis}
In this section, we present an integral-form expression to determine the exact  outage probability, average SNR and ergodic rate. Considering the intractability of the integration,  we use the upper bound approximation of the harmonic mean and derive closed-form analytical bounds  of these parameters. The derived expressions show the system behavior in a relay-assisted environment under the combined effect of pointing errors and fog. 

\subsection{Distribution Function of End-to-End SNR}
We first present  an exact  expression (with an exception of ignoring $1$ in the denominator of \eqref{inst_snr}) for the PDF of end-to-end SNR and then represent a tractable PDF for the performance analysis. Denoting the PDF of $\gamma_1$ and $\gamma_2$  as $f_1(\gamma)$ and $f_2(\gamma_2)$, respectively, we  evaluate the PDFs of $1/\gamma_1$ and $1/\gamma_2$ using the inverse distribution, and employ the PDF of their sum [\cite{Prob_stats} eq.5.55] to get resulting PDF of  $\gamma_{e2e}$ as
\begin{eqnarray}
f(\gamma) = \gamma\int\limits_{t_{min}}^{t_{\rm max}} f_1\left(\frac{\gamma}{t}\right)f_2\left(\frac{\gamma}{1-t}\right)\frac{dt}{t^2(1-t)^2}
\label{exact_pdf}
\end{eqnarray}
Considering  the logarithmic term in \eqref{snr_pdf} positive and applying this constraint to $f_1\left(\frac{\gamma}{t}\right)$ and $f_2\left(\frac{\gamma}{1-t}\right)$, we get
$\ln\left(\frac{A_0^2\gamma_{0}t}{\gamma}\right) > 0 \implies t>\frac{A_0^2\gamma_{0}}{\gamma}= t_{\rm min}$, and 
$
\ln\left(\frac{A_0^2\gamma_{0}(1-t)}{\gamma}\right) > 0 \implies t<1-\frac{A_0^2\gamma_{0}}{\gamma}= t_{\rm max}$.  Note that using these limits,  we can get the exact PDF in \eqref{exact_pdf}. This representation is used to numerically evaluate the OWC performance as a benchmark for our derived analytical expressions.

Since the integral-form expression in \eqref{exact_pdf} is intractable for the PDF in  \eqref{exact_pdf}, we use an upper-bound approximation  $\gamma_{e2e}=\min(\gamma_{1},\gamma_{2})$, to get the cumulative distribution function (CDF) and PDF \cite{Yang2014_relay}:
\begin{eqnarray}
F(\gamma)= F_1(\gamma)+F_2(\gamma)-F_1(\gamma)F_2(\gamma)
\label{aprox_snr_cdf}
\end{eqnarray}
\begin{eqnarray}
f(\gamma)= f_1(\gamma)+f_2(\gamma)-f_1(\gamma)F_2(\gamma)-f_2(\gamma)F_1(\gamma)
\label{aprox_snr_pdf}
\end{eqnarray}
where $f_1(\gamma)$, $f_2(\gamma)$, $F_1(\gamma)$ and $F_2(\gamma)$ can be obtained from  \eqref{snr_pdf} and \eqref{snr_cdf} using  $d_1=d_2=d/2=d_{\rm r}$, where $d_1$ is the distance between source to relay and $d_2$ is the distance from relay to the destination. It is noted that distribution functions in  \eqref{aprox_snr_cdf} and \eqref{aprox_snr_pdf} involves incomplete gamma functions and exponential integrals which requires novel approaches to performance analysis. 

\subsection{Outage Probability}
Outage probability is performance measure to demonstrate the effect of fading channel. It defined as the probability of
failing to reach a specified quality of service (QoS), for example, an SNR threshold value $\gamma_{th}$. Using the PDF in \eqref{exact_pdf}, an exact integral representation  of the outage probability is given as
\begin{eqnarray}
P_{out}= P(\gamma < \gamma_{th}) = \int\limits_{0}^{\gamma_{\rm th}} f(\gamma) d\gamma
\label{outage_prob}
\end{eqnarray}
Further, we can use the CDF in \eqref{aprox_snr_cdf} with $\gamma=\gamma_{\rm th }$ to get  a closed-form approximation of the outage probability. 
	\begin{my_proposition}
	\label{prop_diversity}
		If $k$ and $\beta$ are  the parameters of foggy channel,  $A_0$ and $\rho$ are the parameters of pointing error, and $z_r=4.343/\beta d_r$ with $d=2d_r$ as the transmission link length for a relay-assisted  OWC system, then  an expression for the outage probability is given as
	\begin{eqnarray}
P_{out}=2P_{out}'-(P_{out}')^2,~\text{where}
\label{div_order}
\end{eqnarray}
	\begin{eqnarray}
	&P_{out}'\approx \left(\frac{z_r}{z_r-\rho^2}\right)^k\left(\frac{{A^2_0}\gamma_0}{\gamma_{th}}\right)^{-\rho^2/2}-\nonumber\\&\frac{z_r^k}{\Gamma[k](z_r-\rho^2)}\left(\ln\left(\frac{{A_0}\sqrt{\gamma_0}}{\sqrt{\gamma_{th}}}\right)\right)^{k-1}\left(\frac{{A^2_0}\gamma_0}{\gamma_{th}}\right)^{-z_r/2}\nonumber\\&+\frac{\left(z_r\ln\left(\frac{{A_0}\sqrt{\gamma_0}}{\sqrt{\gamma_{th}}}\right)\right)^{k-1}\left(\frac{{A^2_0}\gamma_0}{\gamma_{th}}\right)^{-z_r/2}}{\Gamma[k]}+\nonumber\\&\frac{(k-1)}{\Gamma[k]}\left(z_r\ln\left(\frac{{A_0}\sqrt{\gamma_0}}{\sqrt{\gamma_{th}}}\right)\right)^{k-2}\left(\frac{{A^2_0}\gamma_0}{\gamma_{th}}\right)^{-z_r/2}
	\label{outage_exact}
	\end{eqnarray}
\end{my_proposition}
\begin{IEEEproof}
Using $\gamma={\gamma_{th}}$, $m=z_r-\rho^2$,  $E_n\left(2-k,z_r\ln u\right)=(z_r\ln u)^{1-k}\Gamma[k-1,z_r\ln u]$ and approximation of incomplete Gamma function $\Gamma[k,m\ln u] \approx u^{-m} (m\ln u)^{k-1}$\cite{Zwillinger2014} in \eqref{snr_cdf}, we get the terms of \eqref{outage_exact}, and thus  \eqref{div_order}. 	
\end{IEEEproof}
It can be seen from the outage probability expression  that  the exponent  of the SNR $A_0^2 \gamma_0$ is $z_r/2$. Thus, the diversity order, $M\approx \frac{2.1715}{\beta d_r}$. 
\subsection{Average SNR and Ergodic Rate}
Exact expressions of the average SNR $\bar{\gamma}^{\rm exact}$ and ergodic rate $\bar{\eta}^{\rm exact}$ are:
\begin{eqnarray*}
\bar{\gamma}^{\rm exact}= \int\limits_{0}^{\gamma_{\rm max}}  \gamma f(\gamma) d\gamma ~\text {and}~
\bar{\eta}^{\rm exact}=\int\limits_{0}^{\gamma_{\rm max}}  \log_2(1+\gamma) f_\gamma(\gamma) d\gamma
\label{siso_rate_exact}
\end{eqnarray*}
where $\gamma_{\rm max}= \frac{A_0^2\gamma_{0}}{2}$ and  $f(\gamma)$  as given in  \eqref{exact_pdf}. Note that we have assumed that relay requires negligible time to relay the data  while computing the ergodic rate.  Considering the intractability of the above integral, we use the PDF in \eqref{aprox_snr_pdf} to derive a closed-form approximations.   Since the fog and pointing error parameters are the same for source to relay and relay to the destination, expressions for the average SNR and ergodic rate  are
\begin{eqnarray}
\label{siso_exact_proof}
& \bar{\gamma}= 2 \int\limits_{0}^{A_0^2\gamma_0}  \gamma \left[f(\gamma)-f(\gamma)F(\gamma) \right]d\gamma\\&
\bar{\eta}= 2 \int\limits_{0}^{A_0^2\gamma_0}  \log_2(1+\gamma) \left[f(\gamma)-f(\gamma)F(\gamma) \right]d\gamma
\label{siso_exact_proof_rate}
\end{eqnarray}

\begin{my_theorem}
	\label{theorem_siso_snr_exact}
	If $k$ and $\beta$ are  the parameters of foggy channel,  $A_0$ and $\rho$ are the parameters of pointing error, and $z_r=4.343/\beta d_r$ with $d=2d_r$ as the transmission link length for a relay-assisted  OWC system, then an approximation of average SNR and ergodic rate are given in \eqref{siso_snr_exact} and \eqref{siso_rate_exact_1}.
	\begin{figure*}
	\begin{eqnarray}
	&\bar{\gamma}\approx A_0^2 \rho^2 z_r^k \gamma_0\bigg[m^{-2 k} \bigg(\frac{\big(1-2\left(\frac{2 + m + 2 \rho^2}{m}\right)^{-k}\big)z_r^k}{1 + \rho^2}+m^k \bigg(-2z_r^{-1+k} (2 + \rho^2 + z_r)^{-k}+\nonumber\\&\frac{-2 - 3 \rho^2 + 2 (1 + \rho^2) \left(\frac{2 + m + \rho^2}{m}\right)^{-k}+ 2 (1 + \rho^2) \left(\frac{2 + \rho^2+z_r}{z_r}\right)^{1-k}}{2 + 3\rho^2+\rho^4}\bigg)\bigg)+\frac{ z_r^{-2 + k} (1 + z_r)^{1 - 2 k} (m+2 m z_r - z_r^2) \Gamma[-1/2 + k]}{m^2\sqrt{\pi} \Gamma[k]}\Big]
	\label{siso_snr_exact}
	\end{eqnarray}
	\end{figure*}
		
	\begin{figure*}	
	\begin{eqnarray}
	&\bar{\eta}\approx\frac{1}{\log 2}\Bigg[2m^{-k}\ \rho^2 z_r^k\bigg(\frac{2(-1+\rho^2\log A_0)}{\rho^4}+2z^{-1+k}(\rho^2+z)^{-1-k}(k-(\rho^2+z)\log A_0)-\nonumber\\&\frac{2}{\rho^4}\bigg(-1+\rho^2\log A_0+\frac{\left(\frac{m+\rho^2}{m}\right)^{-1-k}(m + (1 + k) \rho^2 - \rho^2 (m + \rho^2) \log A_0)}{m}\bigg)+\frac{z_r^k}{m^k\rho^4}\bigg(-1+2\rho^2\log A_0+\nonumber\\&\frac{\left(1+\frac{2\rho^2}{m}\right)^{-k}(m + 2(1 + k) \rho^2 - 2\rho^2 (m + 2\rho^2) \log A_0)}{m+2\rho^2}\bigg)+\frac{2}{\rho^4}\bigg(1-\rho^2\log A_0+\frac{\left(\frac{z+\rho^2}{z}\right)^{-k}(-z- k\rho^2 + \rho^2 (z + \rho^2) \log A_0)}{z}\bigg)+\frac{\log\gamma_0}{\rho^2}+\nonumber\\&\frac{\left(-1+\left(\frac{m+\rho^2}{m}\right)^{-k}\right)\log \gamma_0}{\rho^2}+\frac{\left(1-\left(1+\frac{2\rho^2}{m}\right)^{-k}\right)z_r^k\log \gamma_0}{m^k\rho^2}-z_r^{-1 + k} (\rho^2 + z_r)^{-k} \log\gamma_0-\frac{\left(1-\left(\frac{z+\rho^2}{z}\right)^{1-k}\right)\log \gamma_0}{\rho^2}\bigg)\nonumber\\&-\frac{m^{-2 k} z_r^{2 k} (-1 + \rho^2 (2 \log A_0 + \log \gamma_0))}{\rho^2 \Gamma[k]}-\frac{\rho^2 \Gamma[-\frac{1}{2} + k] ((-3 + 4 k) m + z_r - 2 k z_r + 
		z_r (-2 m + z_r) (2 \log A_0 + \log \gamma_0))}{m^2 \sqrt{\pi} z_r \Gamma[k]}\Bigg]
	\label{siso_rate_exact_1}
	\end{eqnarray}
	\end{figure*}
\end{my_theorem}

\begin{IEEEproof}	
Substituting $u=\frac{A_0}{\sqrt{{\gamma}/{\gamma_0}}}$, $m=(z_r-\rho^2)$ and $E_n\left(2-k,z_r\ln u\right)=\Gamma[k-1,z_r\ln u](z_r\ln u)^{1-k}$ in \eqref{siso_exact_proof}:
		\begin{eqnarray}
	&\bar{\gamma}= \frac{2 A_0^2 z_r^k\rho^2\gamma_0}{\Gamma[k]m^k }\int\limits_{1}^{\infty}\Big[\Big(\frac{\Gamma[k]}{u^{\rho^2+3}}-\frac{\Gamma[k,m\ln u]}{u^{\rho^2+3}}\Big)-\Big(\frac{z_r^k}{m^k} u^{-\rho^2}\nonumber\\&-\frac{z_r^k m^{-k}}{\Gamma[k]}u^{-\rho^2}\Gamma[k,m\ln u]+\frac{u^{-z_r}(z_r\ln u)^{k-1}}{\Gamma[k]}+\nonumber\\&\frac{(k-1)\Gamma[k-1,z_r\ln u]}{\Gamma[k]} \Big)\Big(\frac{\Gamma[k]}{u^{\rho^2+3}}-\frac{\Gamma[k,m\ln u]}{u^{\rho^2+3}}\Big)\Big]du
	\label{siso_snr_exact_proof_1}
	\end{eqnarray}
We use approximation of incomplete Gamma function $\Gamma[k,m\ln u] \approx u^{-m} (m\ln u)^{k-1}$ in  	\eqref{siso_snr_exact_proof_1}, and apply the following identities on the resultant integrals:
	\begin{eqnarray}	\label{eq:identity_1}
&	\int\limits_{1}^{\infty} \left(\ln u\right)^{p} u^{-n} du=\frac{\Gamma[p+1]}{(n-1)^{p+1}}\\&
	\int\limits_{1}^{\infty} u^{-n} \Gamma[k,n\ln u] du=\frac{\left(1-n^k(2n-1)^{-k}\right)\Gamma[k]}{n-1}.
	\label{eq:identity_2}
	\end{eqnarray}
{\color{blue} To get the identity in \eqref{eq:identity_1}, we substitute $\ln u=t$ and apply the definition of Gamma function. Further, the identity in \eqref{eq:identity_2} can be found by substituting $n\ln u=t$  and using  the well known identity  $\int_{0}^{\infty}e^{-at}\Gamma(b,t)dt=a^{-1}\Gamma(b)(1-(a+1)^{-b})$ [\cite{Zwillinger2014}--pp.657].} Using \eqref{eq:identity_1} and 	\eqref{eq:identity_2} in 	\eqref{siso_snr_exact_proof_1} with some algebraic simplifications, we get \eqref{siso_snr_exact} of Theorem \ref{theorem_siso_snr_exact}. Similarly to obtain \eqref{siso_rate_exact_1}, we use the inequality $\log(1+\gamma)\geq\log(\gamma)$  in \eqref{siso_exact_proof_rate} and follow the same procedure used in deriving expression of the average SNR.	
\end{IEEEproof} 
As noted in \cite{Esmail2017_Photonics},  the OWC performance is not sufficient for high-speed data transmission under  moderate ($k>2$) foggy conditions with  pointing error. We derive an exact expressions on the average SNR and ergodic rate for $k=2$ (i.e., under light foggy conditions).

\begin{my_lemma}
	\label{lemma1_siso_exact}
	For light foggy condition $k=2$, expressions for average SNR and ergodic rate are given as:
	\begin{eqnarray}
&	\bar{\gamma}= 2A_0^2 \rho^2 z_r^2\gamma_0\Big(\frac{1}{(2+\rho^2)(2+z_r)^2}-\nonumber\\&\frac{(2 (1 + z_r)^3 + \rho^4 (1 + 2 z_r) + \rho^2 (3 + 4 z_r (2 + z_r)))}{4 (1 + \rho^2) (1 + z_r)^3 (2 + \rho^2 + z_r)^2}\Big)
	\label{siso_exact}
	\end{eqnarray}
	\label{lemma1_siso}
	\begin{eqnarray}
&	\bar{\eta} \geq  2\Big(\big(\frac{\rho^2 z_r (2 \log A_0 + \log \gamma_0)-2 (2 \rho^2 + z_r)}{\rho^2 z_r \log 2}\big)\nonumber \\ &-0.36\big(\rho^2(\rho^2 + z_r)^{-2}-2\rho^{-2} - 5z_r^{-1}  - \nonumber \\&3(\rho^2 + z_r)^{-1}+ 4 \log A_0 + 2 \log \gamma_0\big)\Big)
	\label{siso_rate}
	\end{eqnarray}
\end{my_lemma}
\begin{IEEEproof}
Using $k=2$, and substituting $u=\frac{A_0}{\sqrt{{\gamma}/{\gamma_0}}}$, $m=(z_r-\rho^2)$, and using the series expansion $\Gamma(a,t) \triangleq (a-1)! e^{-t} \sum_{n=0}^{a-1} \frac{t^n}{n!}$ to get $\Gamma(2,m\ln u)=u^{-m}(1+m\ln u)$ and $E_n\left(0,z_r\ln u\right)=\frac{e^{-z_r\ln u}}{z_r\ln u}=\frac{u^{-z_r}}{z_r\ln u}$ in \eqref{siso_exact_proof}:
		\begin{align}
	&\bar{\gamma}= \frac{2 A_0^2 z_r^2\rho^2\gamma_0}{(z_r-\rho^2)^2 }\int\limits_{1}^{\infty}\Big[\Big(u^{-\rho^2-3}-u^{-z_r-3}- (z_r-\rho^2)\nonumber \\&u^{-z_r-3}\ln u\Big) - \Big(z_r^2(z_r-\rho^2)^{-2}u^{-\rho^2}-z_r^2(z_r-\rho^2)^{-2}u^{-z_r}\nonumber \\&-z_r^2(z_r-\rho^2)^{-1}u^{-z_r}\ln u+u^{-z_r}+z_r u^{-z_r}\ln u \Big)\nonumber \\& \big(u^{-\rho^2-3}-u^{-z_r-3}-(z_r-\rho^2)u^{-z_r-3}\ln u\big)\Big]du.
	\label{siso_exact_proof_1}
	\end{align}
In order to find closed form expressions of the above integrals, we use the identity as given in \eqref{eq:identity_1}. Thus,  applying	\eqref{eq:identity_1} in \eqref{siso_exact_proof_1}, we  get  \eqref{siso_exact} of Lemma \ref{lemma1_siso_exact}. Similarly, in order to obtain \eqref{siso_rate}, we use the inequality $\log(1+\gamma)\geq\log(\gamma)$ in \eqref{siso_exact_proof_rate} and follow the same procedure used in deriving expression of the average SNR.	

\end{IEEEproof} 

\begin{figure*}[t]
	\begin{center}
		\subfigure[$P_{\rm out}$ versus $d_r$ at $P_t = 15$ \mbox{dBm} and $d = 1$ \mbox{km}.]{\includegraphics[width=\columnwidth]{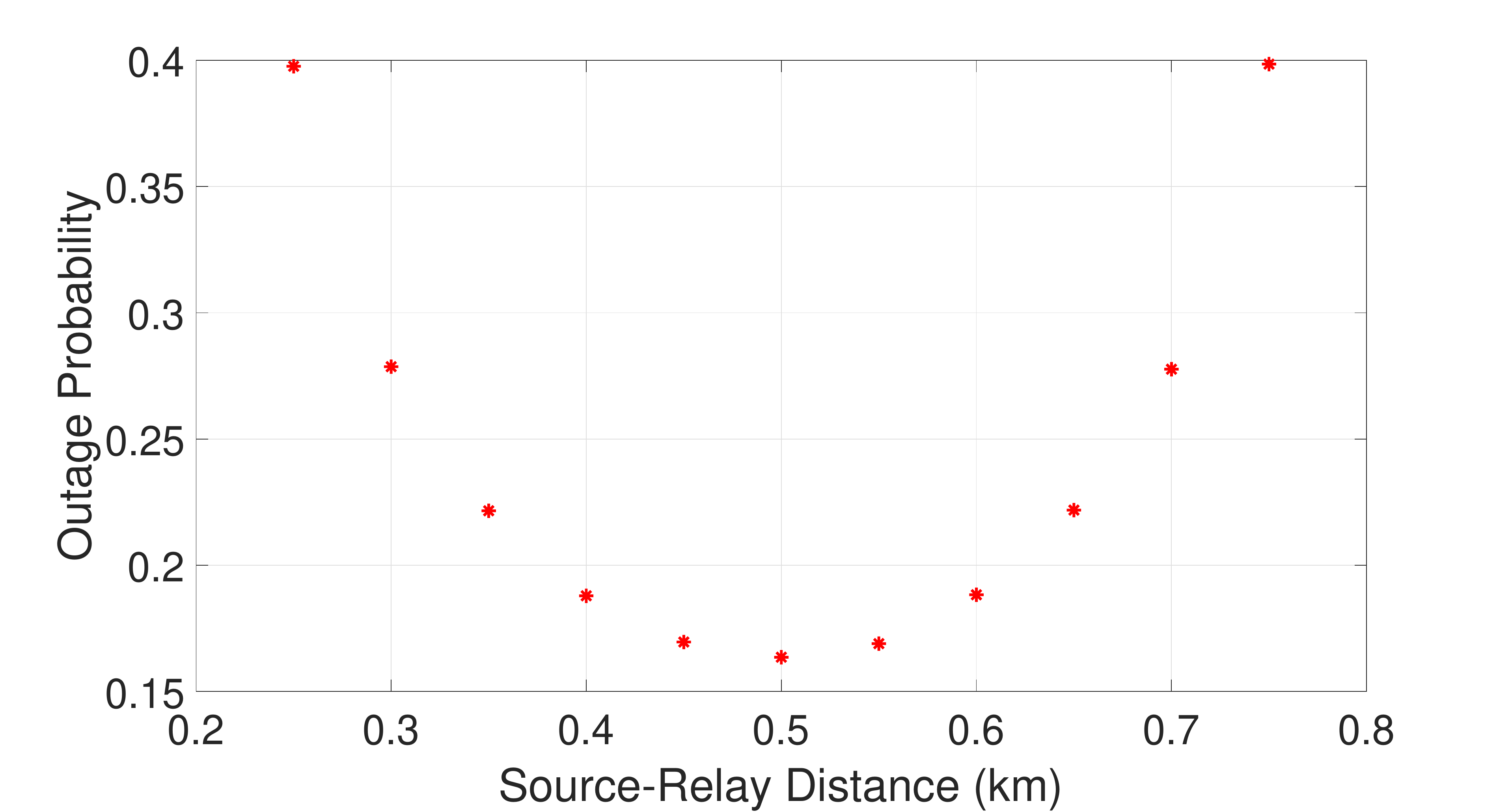}}
		\subfigure[Outage probability.]{\includegraphics[width=\columnwidth]{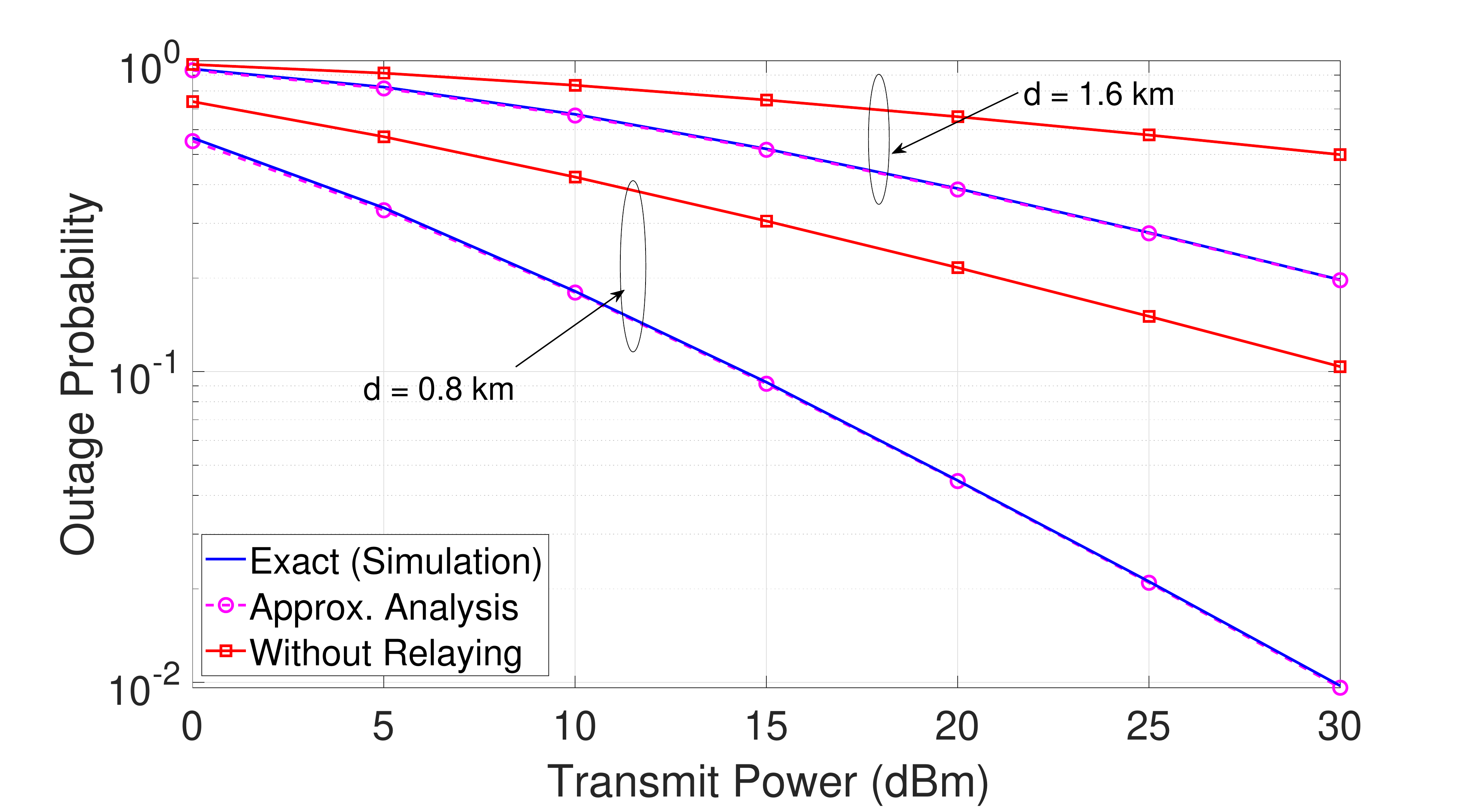}}
		\caption{Outage performance of relay-assisted OWC system under foggy channel with pointing error.}
		\label{out_perf}
	\end{center}
\end{figure*}

Note that the first term in \eqref{siso_exact} and \eqref{siso_rate} with $d_r=d$ corresponds to twice of  the average SNR and ergodic rate  without relaying. Thus, $\bar{\gamma}_{\rm direct}= \frac{z^2A_0^2 \rho^2 \gamma_0}{(2+\rho^2)(2+z)^2}$ \cite{rahman2020cl}. Since all the terms in \eqref{siso_exact} is positive, we expect a higher average SNR with relaying. This has been extensively studied through numerical analysis in the following section.

\begin{figure*}[t]
	\begin{center}
		\subfigure[Average SNR.]{\includegraphics[width=\columnwidth]{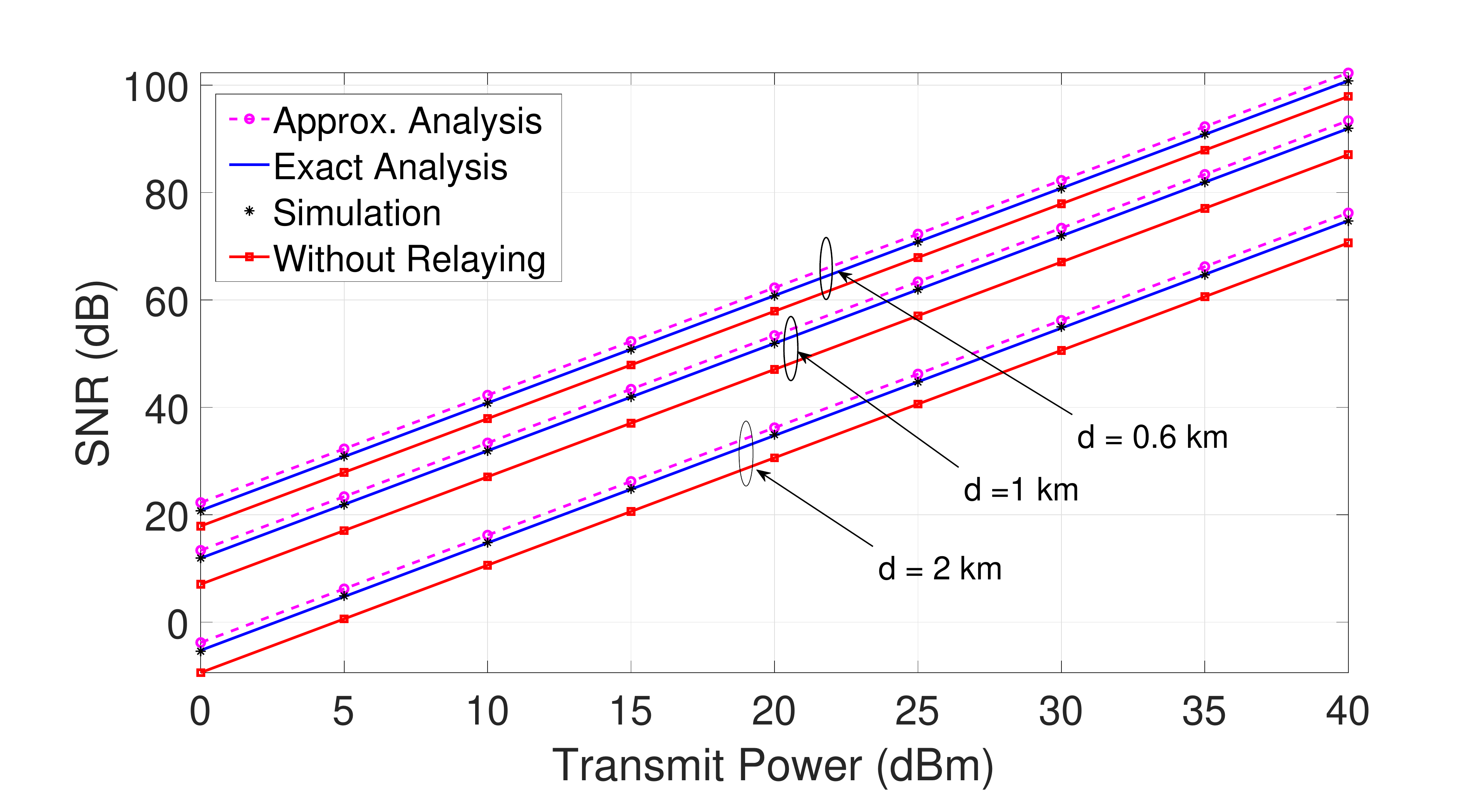}}
		\subfigure[Ergodic rate.]{\includegraphics[width=\columnwidth]{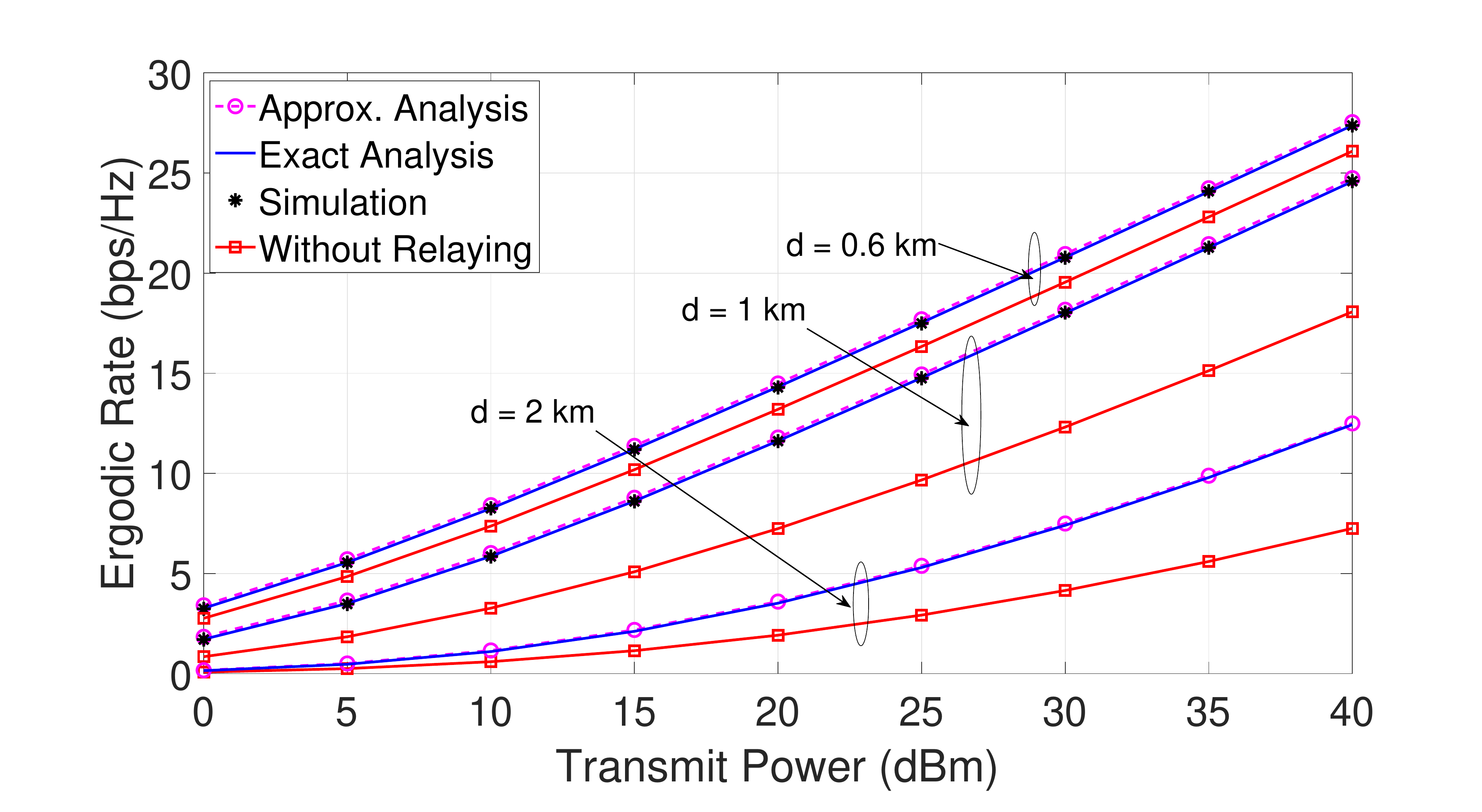}}
		\caption{Average SNR and ergodic rate performance of relay-assisted OWC system under foggy channel with pointing error.}
		\label{snr_perf}
	\end{center}
\end{figure*}

\section{Numerical Analysis}
In this section, we use numerical analysis and Monte Carlo simulation (averaged over $10^6$ channel realizations) to demonstrate the outage probability, average SNR, and ergodic rate performance of the relay-assisted OWC system  under the combined effect of fog and  pointing error. We provide a comparison between the performance of direct and relay-assisted transmissions at various link distances. We use the light foggy conditions and consider the experimental values of foggy channel  parameters ($k=2$ and $\beta=13.12$) \cite{Esmail2017_Access}.  We consider the range for transmitted power as $P_t=0$ \mbox{dBm} to $30$ \mbox{dBm}, receiver with AWGN power $\sigma_w^2=10^{-14}$, and detector responsivity $R=0.41$ \mbox{A/W}.

   Since the experimental values of  pointing error parameters $\rho$ and $A_0$ are available for $1$ \mbox{km} \cite{Esmail2016_Photonics}, we use Gaussian beam optics to determine these parameters for other link distances for a better estimate of performance. The radius of a Gaussian beam $w_{z}$ (i.e., the beam waist) increases non-linearly with distance near the focal point of the transmitter but after the Rayleigh distance $\frac{\pi w_0^2}{\lambda}$, the beam waist increases linearly with the distance.  Here, $w_0$ is the waist size at the focal point, and $\lambda$ is the light wavelength. For a helium neon laser ($\lambda = 650$\mbox{nm}) with $w_0 = 5$ \mbox{mm}, we get Rayleigh distance of $50m$. Thus, we can use the linear approximation $w_{z} = \theta d$, where $\theta$ is beam divergence. Using  $w_z=2.5m$ at $d = 1$ \mbox{km} \cite{Farid2007}, we get $w_{z} = \frac{2.5d}{1000}$, and thus  $\upsilon=\sqrt{\pi/2}\ a/\omega_z=\sqrt{\frac{\pi}{2}}\frac{1000a}{2.5d}$. Finally, using $w_{zeq}= w_z \text{erf}(\upsilon)/(2\upsilon \exp(-\upsilon^2))$, we can get  $A_0=(\mbox{erf}(\upsilon))^2$ and $\rho=\omega_{z_{eq}}/2\sigma_s$, where $\sigma_s=0.28$ \mbox{m} \cite{Farid2007}. The parameter $\sigma_s$  is caused by  building swaying and thermal expansion and it is taken to be constant with distance.

First,  we analyze the optimal location of relay by  changing the source-relay distance from $250$ \mbox{m} to $750$ \mbox{m} for an OWC system with the source-to-destination distance of $1$ \mbox{km}. It can be seen from Fig.~\ref{out_perf}a that the  minimum outage performance occurs at $d_r = 500$ \mbox{m}. Thus, we see that the symmetric analysis considered in the paper  is justified giving a near-optimal performance for the OWC system. Note that the similar conclusion holds when we analyze the optimal relay location using average SNR and ergodic rate.

 In Fig.~\ref{out_perf}b, we demonstrate the diversity order of system by analyzing  the outage probability   performance of the OWC system  at  two transmission link lengths ($d=0.8$ \mbox{km} and $d=1.6$ \mbox{km}) by considering the threshold SNR $\gamma_{th} = 6$ \mbox{dB}. The figures shows that the relay-assisted channel provides a significant gain in the outage performance comparing the single-link channel. It can be seen that the direct transmission requires almost $30$ times of transmit power to achieve the same outage probability  (i.e., $P_{\rm out}=0.1$ at $30$ \mbox{dBm}) with relaying at a transmit power of  $15$ \mbox{dBm}. It can also be seen that decreasing the link distance by half doubles the diversity order of OWC system, as predicted through analysis in subsection III-B. It can also be seen that derived approximation of \eqref{outage_prob} excellently matches with the simulation results.

In  Fig.~\ref{snr_perf}a, we plot the average SNR at different transmitted powers for three transmission link lengths ($d=0.6$ \mbox{km}, $d=1.0$ \mbox{km} and $d=2.0$ \mbox{km}).  As expected, the performance degradation with link length is  evident here, being $\approx9$ \mbox{dB} from $0.6$ \mbox{km} to $1$ \mbox{km}, and $\approx17$ \mbox{dB} from $d=1$ \mbox{km} to $d=2$ \mbox{km}. It can  be seen that  relay and single-link SNRs differ by $4.7$ \mbox{dB} (approximately $3$ times in linear scale), which is a significant improvement in the SNR performance.  However,  compared to the average SNR,  an enhanced performance improvement for ergodic rate with relaying can be observed in Fig.~\ref{snr_perf}b. We also notice the distance-dependent benefit in ergodic rate performance for the relay-assisted channel compared to direct transmission. It can be seen that gain with relaying is larger for $d=1$ \mbox{km} than $d=0.6$ \mbox{km} and $d=2$ \mbox{km}. It is noted that we have assumed that relay consumes negligible time to relay the data. It can  also be seen that derived approximation for average SNR and ergodic rate are very close comparing  the exact numerical  analysis and Monte Carlo simulations.

\section{Conclusions}
In this paper, we investigated the performance of an AF relaying based OWC system under the combined effect of fog and pointing error. We provided a detailed analysis for outage probability, average SNR, and ergodic capacity in order to  show the benefit of  proposed dual-hop relaying for the OWC system with weak direct links. Numerical analysis and Monte Carlo simulations show that the derived analytical bounds are  close to the complicated integral expressions, and thus  can be implemented for  real-time tuning of the system parameters for optimized performance.   It was also demonstrated that the relay-assisted system shows better performance than the single-link transmissions. The relaying scheme requires almost $30$ times less transmission power to achieve the same outage probability  and  achieves an SNR gain of approximately $4.7$ \mbox{dB} compared to the direct transmission. The extended version of the work will involve a consideration of asymmetric situations with different channel statistics at the relay and destination.

		\bibliographystyle{ieeetran}
		\bibliography{IEEE_CL_bibfile_final}	
		
\end{document}